\documentclass{article}
\usepackage{spconf,amsmath,graphicx}
\usepackage{booktabs}
\usepackage{amsmath}
\usepackage{graphicx}
\usepackage{amsmath,amssymb,amsfonts}
\usepackage{bm}
\usepackage{epsfig,graphicx,subfigure}
\usepackage{color}
\usepackage[normalem]{ulem}
\usepackage{multirow}
\usepackage{booktabs}
\usepackage{multirow}
\usepackage[lowtilde]{url}
\usepackage[hidelinks]{hyperref}
\usepackage{cite}
\usepackage[ruled,vlined]{algorithm2e}
\usepackage{bm}

\makeatletter
\def\blfootnote{\gdef\@thefnmark{}\@footnotetext}
\makeatother

\title{TFGAN: Time and Frequency Domain Based Generative Adversarial Network for High-fidelity Speech Synthesis}
\name{Qiao Tian$^1$, Yi Chen$^2$, Zewang Zhang$^1$, Heng Lu$^1$, Linghui Chen$^1$, Lei Xie$^2$, Shan Liu$^1$}

\address{$^1$Tencent, China \\$^2$Audio, Speech and Language Processing Group (ASLP@NPU), School of Computer Science, \\ Northwestern Polytechnical University, Xian, China\\
  {\small \tt \{briantian, zewangzhang, bearlu, nedchen, shiningliu\}@tencent.com}
  \\
  {\small \tt \{yichen, lxie\}@nwpu-aslp.org}}

\begin{document}
%
\maketitle
\begin{abstract}
Recently, GAN based speech synthesis methods, such as MelGAN, have become very popular. 
Compared to conventional autoregressive based methods, parallel structures based generators make waveform generation process fast and stable.
However, the quality of generated speech by autoregressive based neural vocoders, such as WaveRNN, is still higher than GAN.
To address this issue, we propose a novel vocoder model: TFGAN, which is adversarially learned both in time and frequency domain.
On one hand, we propose to discriminate ground-truth waveform from synthetic one in frequency domain for offering more consistency guarantees instead of only in time domain.
On the other hand, in contrast to the conventionally frequency-domain STFT loss approach or feature map loss by discriminator to learn waveform, we propose a set of time-domain loss that encourage the generator to capture the waveform directly.
TFGAN has nearly same synthesis speed as MelGAN, but the fidelity is significantly improved by our novel learning method. 
In our experiments, TFGAN shows the ability to achieve comparable mean opinion score (MOS) than autoregressive vocoder under speech synthesis context.

\end{abstract}
\section{Introduction}
\label{sec:intro}
Neural network-based vocoders have been rapidly improved in recent years, which have obvious advantages over the traditional parametric vocoders~\cite{morise2016world, kawahara2006straight} in the aspect of naturalness of speech. 
Since the emergence of WaveNet~\cite{van2016wavenet}, the neural vocoder has developed dramatically, whose quality is closer to human voice. 
However, due to autoregressive modeling method and very deep structures, the generation efficiency of WaveNet is inherently low. 
To this end, in ~\cite{kalchbrenner2018efficient}, a model named WaveRNN has been proposed, which models waveform with single GRU layer and has much smaller complexity than WaveNet.
Also, LPCNet~\cite{valin2019lpcnet} greatly increased the synthesis speed on the premise of guaranteeing voice quality by combining linear prediction coding with neural vocoder. 
What's more, many signal processing technique has been introduced in neural vocoder for smaller complexity,
such as Multi-band WaveRNN~\cite{yu2019durian} and FeatherWave~\cite{tian2020featherwave}. 
Obviously, autoregressive vocoders are easy to get good quality, but engineered optimizations are required for achieving corresponding inference speed.
Parallel WaveNet~\cite{oord2017parallel} and ClariNet~\cite{ping2018clarinet}) made steps to generate waveform in parallel with inverse normalizing flow~\cite{kingma2016improved}, which required to distill from a autoregressive WaveNet model.
Other flow-based models, such as WaveGlow~\cite{prenger2019waveglow} and  WaveFlow~\cite{ping2019waveflow}, removed the distillation and trained with maximum likelihood directly, but the quality is still not good enough.
Parallel WaveGAN~\cite{yamamoto2020parallel} and MelGAN~\cite{kumar2019melgan} adopt non-autoregressive architecture and adversarial training, 
they can generate waveform in parallel at inference stage. 
Therefore, the synthesis speed is greatly accelerated without extra engineered work. 

Although the synthesis speed of GAN based model is quite charming, the generated quality of speech synthesis is not satisfactory. 
Currently, there is no efficient time-domain loss for generator and discriminator only model the audio in time domain.
As a result, the synthesized audio has obvious artificial traces in the high frequency and there are discontinuous phenomena in the phase, which lead to the problems of metallic sense and lag in hearing sense, such as in MelGAN.
What's more, the widely adopt upsampling mechanism with transpose convolution could introduce artifacts in breathing part of speech. 
In this paper, we propose the TFGAN vocoder based on MelGAN, which encourage generator and discriminator learn waveform both in time and frequency domain.

We summarize the contributions of proposed TFGAN as follows:
Firstly, we propose a ResNet18~\cite{he2016deep} based frequency discriminator to improve the reality of phase and offer more consistency guarantees compared to only by time-domain discriminator.
Secondly, multiple time-domain waveform loss are proposed to help generator to learn directly and reduce metallic noise caused by overdependence of STFT Loss.

\section{Related Work}
\label{sec:format}

\subsection{MelGAN}
\label{ssec:melgan}

 In basic MelGAN, a stack of transposed convolutional layers are used to upsample the mel-spectrogram to match the waveform sequence and each transposed convolutional layer is followed by a stack of residual blocks with dilated convolutions whose receptive field increased exponentially with the number of layers. 
 To solve the metallic sound problem, discriminator adopt a multi-scale architecture which have identical network structure but operate on different audio scales. 
 Multiple discriminator at different scales are motivated from the fact that audio has structure at different levels.
 Specifically, with K discriminators, basic MelGAN conducts adversarial training with objectives as:
 
\begin{equation}
\begin{split}
    D_{loss} = \min_{D_k} \mathbb{E}_x \left [\min(0, 1 - D_k(x)) \right]\\
    + \mathbb{E}_{s,z} [\min(0, 1 + D_k(G(s,z)))], {\forall}k = 1,2,3,
\end{split}
\label{eq:melgan1}
\end{equation}

\begin{equation}
    G_{loss} = \min_{G} \mathbb{E}_{s,z}[ \sum_{k=1}^{3}-D_k(G(s,z))],
\label{eq:melgan2}
\end{equation}
where $x$ represents the raw waveform, $s$ represents the acoustic features(eg., mel-spectrogram) and $z$ represents the gaussian noise vector.

In addition to the discriminator's signal, the feature matching objective is also been used to train the generator. This objective minimizes the L1 distance between the discriminator feature maps of real and synthesized audio.




\subsection{Multi-resolution STFT Loss}
As proposed in~\cite{yamamoto2020parallel, yang2020multi, yang2020vocgan}, STFT Loss can help model generate higher quality speech and fast convergence.
The single STFT loss contains $L_{sc}$ and $L_{mag}$ which denote spectral convergence and log STFT magnitude loss respectively, they are defined as follows: 
\label{ssec:mr_stft}
\begin{equation}
    L_{sc}(x, \widetilde{x})= \frac{\left \|  \left |STFT(x) \right | - \left | STFT(\widetilde{x})\right |\right \|_{F}} {\left \| \left | STFT(x) \right |\right \|_{F} },
\label{eq:loss_sc}
\end{equation}

\begin{equation}
    L_{mag}(x, \widetilde{x})= \frac{1}{N} \left \| log\left | STFT(x)\right | -log\left | STFT(\widetilde{x})\right | \right \|_{1},
\label{eq:loss_mag}
\end{equation}
where $\left \| \cdot \right \|_F$ and $\left \| \cdot \right \|_1$ represent the Frobenius and $L_1$ norms, respectively. $\left |STFT(\cdot) \right |$ indicates the $STFT$ function to compute magnitudes and $N$ is the number of elements in the magnitude.
The multi-resolution STFT loss~\cite{yamamoto2020parallel} is the sum of M single STFT losses with different analysis parameters and the loss($L_{stft}$) is represented as follows:
\begin{equation}
    L_{stft}(G)=\mathbb{E}_{x, \widetilde{x}} [ \frac{1}{M}\sum_{m=1}^{M}\left ( L_{sc}^{m}(x, \widetilde{x}) + L_{mag}^{m}\left ( x,\widetilde{x}\right )\right ) ].
\label{eq:L-mr-stft}
\end{equation}
 
\section{TFGAN}
\label{sec:TFGAN}

\begin{figure}[htp]
    \centering
    \includegraphics[width=0.5\textwidth]{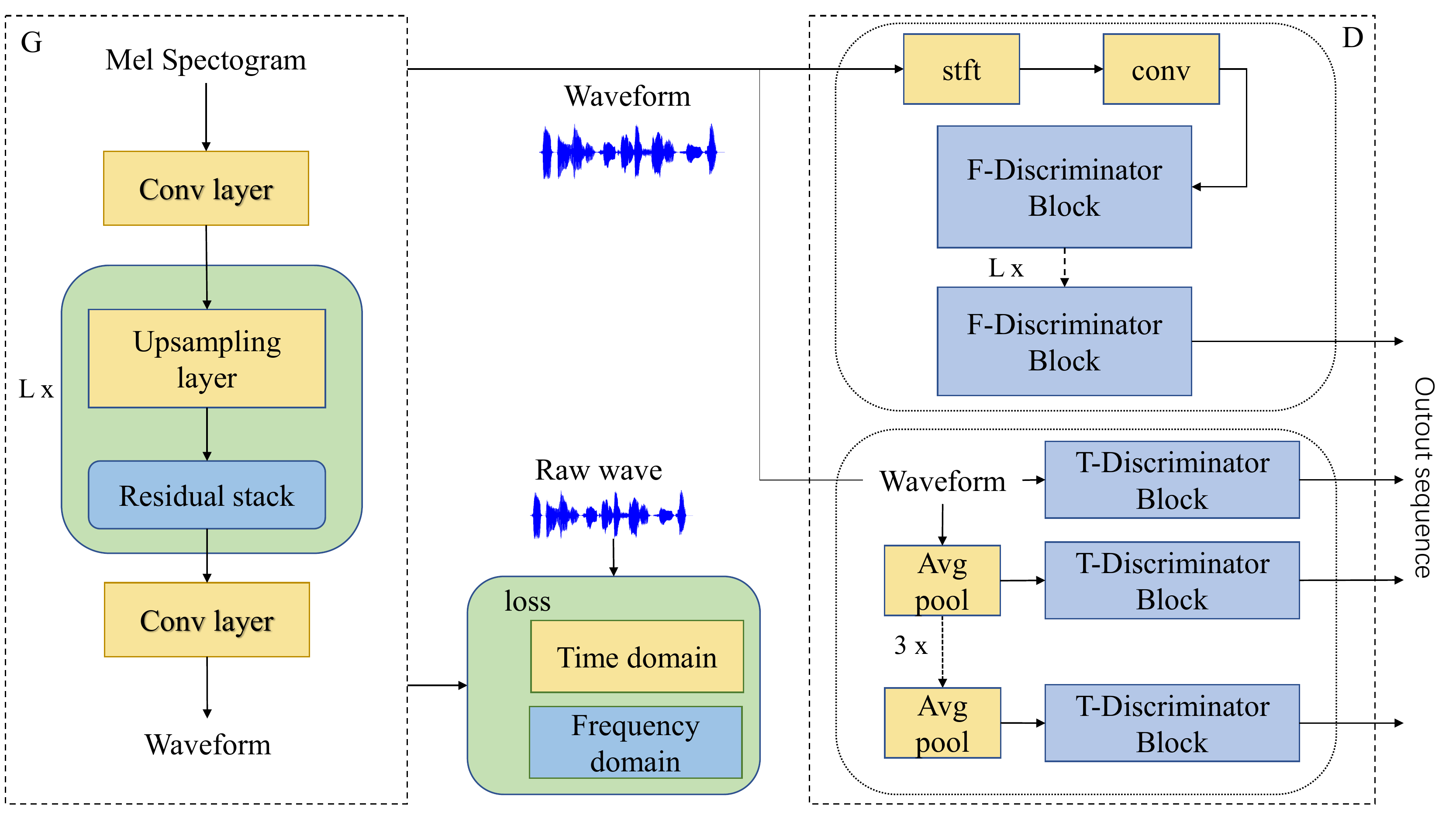}
    \caption{The architecture of proposed TFGAN}
    \label{fig:TFGAN}
\end{figure}

In order to figure out the artificial sound appear in MelGAN\cite{kumar2019melgan} and MB-MelGAN\cite{yang2020multi}, we model speech in both time and frequency domain.
Fig.~\ref{fig:TFGAN} illustrates the proposed TFGAN. The model consists of three trainable components: a generator $G$ and two discriminators: time discriminator and frequency discriminator.

\begin{figure}[htp]
    \centering
    \includegraphics[width=0.5\textwidth]{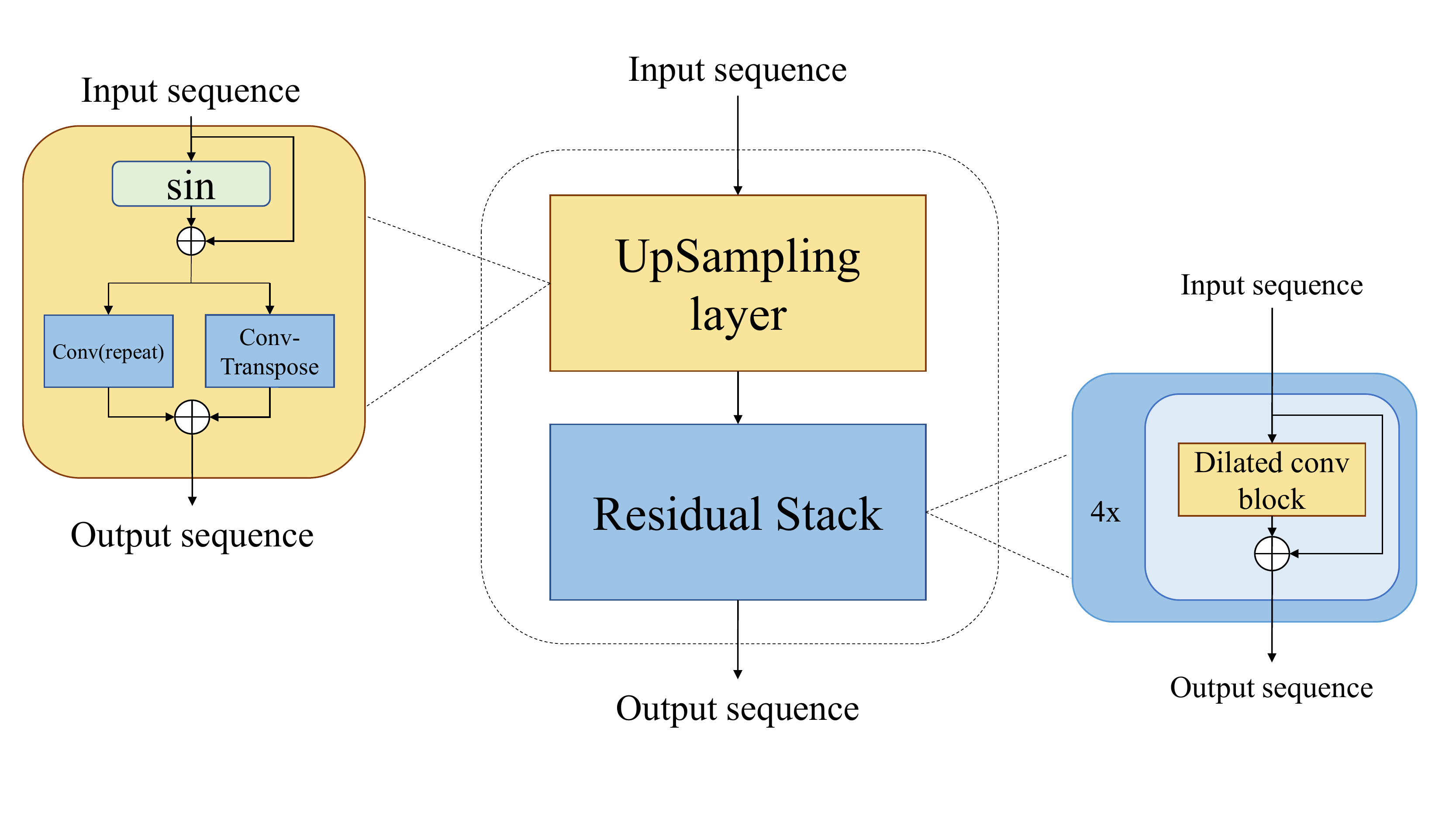}
    \caption{The UpSample and ResStack Block}
    \label{fig:UpSample_block}
\end{figure}

 Similarly to basic MelGAN we also use s stack of transpose convolutional layers to upsample the mel-spectrogram, but in order to remove periodic artifacts in breathing part of speech, we improve the upsampling structure as shown in Fig.~\ref{fig:UpSample_block}.
 Firstly, we feed the input sequence into a sinusoidal activation function, the results of which add to themselves.
 Then, in addition to using transpose convolution, we also repeat the output of the first step by the up-sample factor directly and following a convolutional layer.
 Finally the outputs of transpose convolutional layer and repeat structure are added as the up-sample block output. And 240x upsampling is conducted through 3 upsampling layers with 8x, 6x and 5x factors respectively. 
 Expanding the receptive field appropriately is helpful to improve the quality of speech generation~\cite{yang2020multi}, we increase the depth of the residual dilated convolution stack from 3 layers to 4 layers with dilation 1, 3, 9, 27.

\subsection{Frequency Discriminator}
\label{sec:fd}

In order to generate high quality speech without the issues(e.g., vibrations) occurred with single time domain discriminator, we propose the Frequency Discriminator to correct the output of generator both in the time domain and frequency domain.

In frequency discriminator, we use the the short-time Fourier transform(STFT) to get the characteristics of the audio in the frequency domain, then the real and imaginary parts are fed into four F-Discriminators ($D^{(F)}_{1}, D^{(F)}_{2}, D^{(F)}_{3}, D^{(F)}_{4}$) respectively. Each F-Discriminator consists of two Residual Blocks, each of which contains a stack of convolutional layers. In order to avoid the problem of gradient disappearance caused by too many convolutional layers, $D_{1}^{(F)}$ contains only two convolution layers, but $D_{2}^{(F)}$, $D_{3}^{(F)}$, $D_{4}^{(F)}$ have a residual connection. Basically, we adopt ResNet18~\cite{he2016deep} as frequency discriminator.


The time domain discriminator adopt a multi-scale architecture operating on different audio scales which is similar to that in basic MelGAN: D1 operates on the scale of raw audio, while D2 and D3 operate on raw audio down-sampled by a factor of 2 and 4 respectively.

\begin{figure}[htp]
    \centering
    \includegraphics[width=0.5\textwidth]{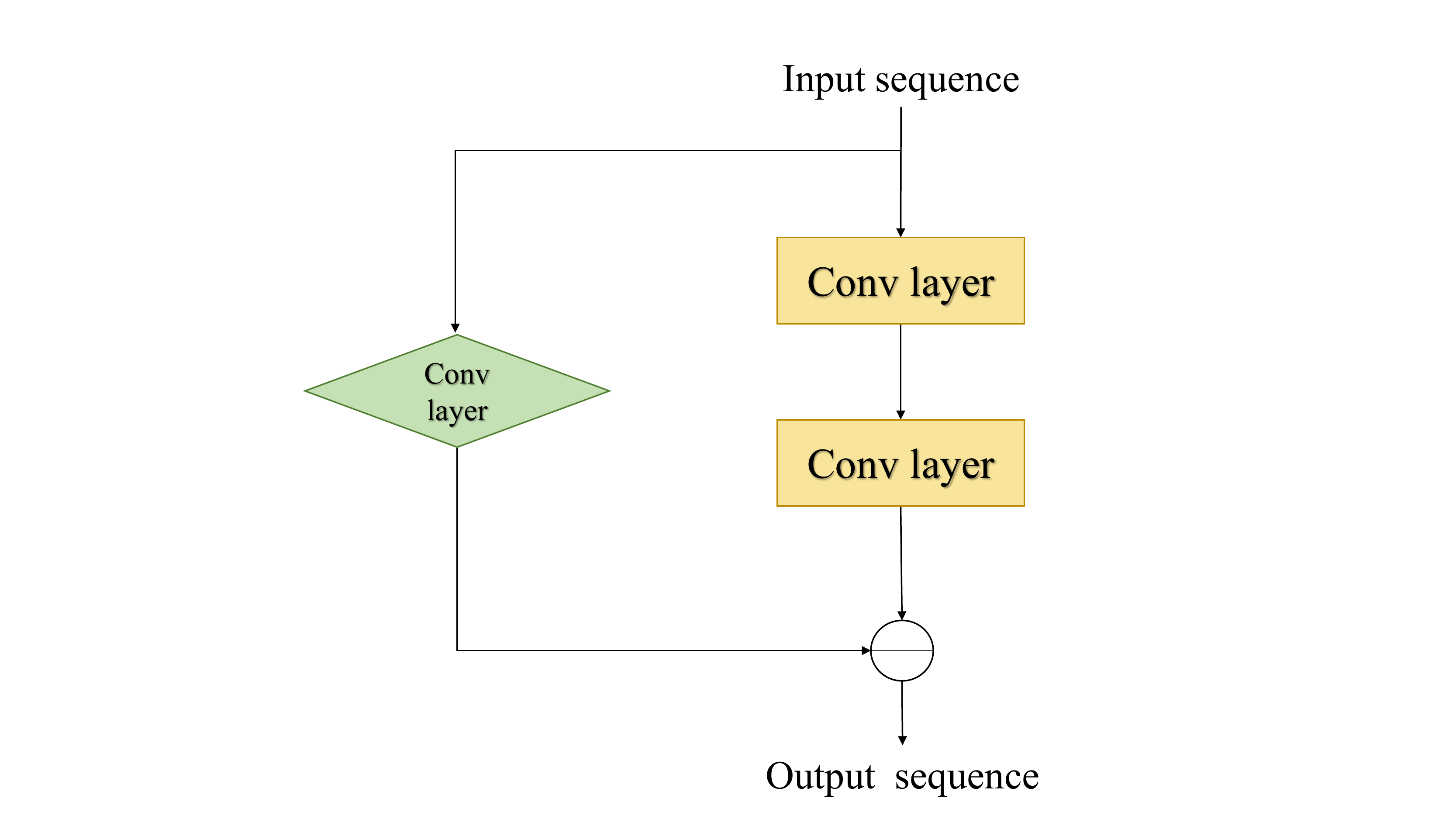}
    \caption{The Residual Block of F-Discriminator}
    \label{fig:residual_block}
\end{figure}

\begin{table}[]
\centering
\caption{The parameters for Multi-resolution audio loss in time domain.}
\begin{tabular}{l|l|l|l|l}
\hline
             & 1 & 2   & 3   & 4   \\ \hline
frame length & 1 & 240 & 480 & 960 \\ \hline
hop length   & 1 & 120 & 240 & 480 \\ \hline
\end{tabular}
\label{tb:dy-loss-param}
\end{table}

\subsection{Time domain audio loss}
\label{ssec:audio_losses}

We find that although the multi-resolution STFT loss helps with convergence, but its periodicity makes the synthesized speech sound artificial in high frequency.
In order to solve this problem, we design a set of loss functions in time domain called Multi-scale dynamic loss, which can help model learn audio in time domain directly. Multi-scale dynamic loss consists of three parts in four resolution scales: $loss_e$, $loss_p$ and $loss_t$ which are designed to capture energy, fast convergence and remove high-frequency metallic noise respectively.

\noindent $a)$~\textbf{Energy loss}:
\begin{equation}
    loss_e = \left \| \mathbb{E}{(x^{2})} - \mathbb{E}{( \widetilde {x}^{2})} \right \|_{1},
\label{eq:energy_loss}
      \end{equation}
where $ \left \| \cdot \right \|_{1}$ is the $L^{1}$ Frobenius norm, $x$ and $\widetilde{x}$ represent the target audio signal and synthetic signal which are processed by frames with window size and hop size showed in Table~\ref{tb:dy-loss-param}.

\noindent $b)$~\textbf{Time loss}:
\begin{equation}
    loss_t = \left \| \mathbb{E}{(x)} - \mathbb{E}{(\widetilde{x})} \right \|_{1},
\label{eq:time_loss}
\end{equation}
where $x$ and $\widetilde{x}$ represent the target audio signal and synthetic signal processed by frames with window size and hop size showed in Table~\ref{tb:dy-loss-param};

\noindent $c)$~\textbf{Phase loss}:
\begin{equation}
    loss_p = \left \| \Delta x - \Delta \widetilde{x}\right \|_{1},
\label{eq:phase_loss}
\end{equation}

where $\Delta x$ represents the signal $x$ processed by the method of first difference. Due to the addition of a frequency domain discriminator, the final objective becomes:

\begin{equation}
 L^{T}(G) = loss_e + loss_t + loss_p,
\label{GLoss_time}
\end{equation}

\begin{equation}
\begin{split}
    \min_{G} \mathbb{E}_{s,z} [\lambda_1 \sum_{k=1}^{3} -D_k(G(s,z))] + \lambda_2 \mathbb{E}_{s}[L_{stft}(G)]\\
    + \lambda_3 \mathbb{E}_{s,z}^{f}[-D( stft(G(s, z)), x)] + \lambda_4 \sum_{k=1}^{4} \mathbb{E}_{s}[L_{k}^{T}(G)].
\label{eq:generator_loss}
\end{split}
\end{equation}

As shown in Eq.~\ref{eq:generator_loss}, We also adopt the multi-resolution STFT loss to speed up the adversarial training and stability as an auxiliary loss.

\section{Experiments}
\label{sec:experiments}
\subsection{Data Set}
\label{ssec:exp_setup}
In our experiments, the training and testing data we used are ground truth aligned(GTA) data from a well-trained Tacotron2~\cite{shen2018natural}. And the origin data is a Mandarin corpus of 20 hours of recordings, which were recorded by a professional broadcaster. All the recordings were down-sampled to 24 kHz sampling rate with 16-bit format. About 19 hours of recordings were used for model training and the rest were used for validation. We set $\lambda_1$, $\lambda_2$, $\lambda_3$ as 1.0 and set $\lambda_4$ as 20.0 specially.

\subsection{Experimental Setup}
\label{ssec:comparison_models}
To demonstrate that the proposed model can synthesize high-fidelity audio, we chose MelGAN and an autoregressive vocoder, Multi-band WaveRNN~\cite{yu2019durian} as the baseline. In the MelGAN baseline system, 
we used the open-source implementation\footnote{https://github.com/kan-bayashi/ParallelWaveGAN} and the configuration was exactly the same as its original version.
For WaveRNN we use 4 bands, 10-bit $\mu$-law quantization for dual softmax layers and the dimension of FC layer is 256~\cite{tian2020featherwave}, the preemphasis parameter is 0.85.

\textbf{Generator} In our proposed TFGAN, we use three upsampling and residual stack combinations to realize $240$x upsampling, where the up-factor is (8, 6, 5). The output channels of the first convolution is 512 and the upsampling network are (256, 128, 64).
The transposed convolution's kernel-size in each upsampling block is twice of the stride, which in the convolution after the repeating layer is 1.

\textbf{Discriminator} We adopt ResNet18~\cite{he2016deep} as the frequency discriminator, which performs in frequency domain.
The waveform passes through a STFT using the hanning window whose fft size, hop length and window length are (512, 240, 512), then the real and imaginary parts of the frame-level results obtained by the STFT are fed into the frequency discriminator respectively. The time domain discriminator is the same as that in Multi-Band MelGAN~\cite{yang2020multi}, each time domain discriminator block has 3 strided convolution with stride 4.

\textbf{Training and inference} In the training phase, the Adam \cite{kingma2014adam} optimizer was adopted with a learning rate of 0.0002 for both generator and discriminator. The proposed model was trained on a single GPU with mini-batch size of 16 audio clips(24000 samples). The weights of the neural vocoders were randomly initialized with fixed random seed and all the networks were trained with 2000k iterations. 

\begin{table}
\centering
\caption{Mean Opinion Score (MOS) with $95\%$ confidence intervals for different models.}
\begin{tabular}{cc}
\toprule
 & {\textbf{MOS on speech quality}} \\
\midrule
MelGAN & 3.95 $\pm$ 0.05 \\
WaveRNN & 4.34 $\pm$ 0.04 \\
\textbf{TFGAN} & \textbf{4.35 $\pm$ 0.04} \\
Ground Truth & 4.42 $\pm$ 0.03 \\
\bottomrule
\end{tabular}
\label{table:mos}
\end{table}

\subsection{Ablation study}
\label{ssec:ablation}
This experiment was an ablation test on TFGAN. In order to avoid indistinguishable error caused by too many audio samples, we use Short-Time Objective Intelligibility(STOI) and Perceptual Evaluation of Speech Quality(PESQ) as objective indicators to evaluate the model~\cite{adiga2018use}.
We conducted four groups of ablation experiments, and the experimental results are shown in Table~\ref{table:STOI-PESQ}.
All the models were trained to 2 million steps. On the basis of $B_0$, Multi-resolution STFT Loss and the modified upsampling structure were added in $P_1$ and $P_2$ respectively, which all improved on the metric PESQ. Periodic artifacts were removed in breathing part with the modification of upsampling layer, especially under TTS context. $P_3$ is an improved combination of $P_1$ and $P_2$ without the addition of a frequency discriminator, with overall improvement in STOI and PESQ, where PESQ surpassed WaveRNN. $P_4$ is our proposed model, which achieved the best in both indices and exceeded Multi-band WaveRNN.

\subsection{Comparison on MOS}
\label{ssec:mos_test}
Subjective evaluation was conducted to evaluate the perceptual quality of the proposed TFGAN vocoder by MOS. 
In order to perform fair comparison, we randomly selected 40 utterance from test set for MOS testing and 30 native Mandarin speakers participated in the listening test.
The results\footnote{https://wavecoder.github.io/TFGAN/} of the subjective MOS evaluation is presented in Table~\ref{table:mos}.The result show that the proposed TFGAN can generate high quality 24 kHz speech with a slightly better MOS than MelGAN and Multi-band WaveRNN.

\begin{table}
\centering
\caption{The objective result of ablation study with STOI and PESQ. The higher is better.}
\begin{tabular}{cccc}
\toprule
  & & {\textbf{STOI}} & {\textbf{PESQ}} \\
\midrule
$B_0$ & Baseline(MelGAN) & 0.94  & 2.77 \\
\midrule
$P_1$ & + STFT loss & 0.94 & 2.92 \\
$P_2$ & \hskip  1em + Residual and sine activation & 0.94 & 2.93 \\
\midrule
$P_3$ & TFGAN w/o F-Discriminator & 0.95 & 3.14 \\
$P_4$ & \textbf{TFGAN} & \textbf{0.95}  & \textbf{3.24} \\
$P_5$ & WaveRNN & 0.92 & 3.02 \\
\midrule
$P_6$ & Ground Truth & 1.00 & 4.50 \\
\bottomrule
\end{tabular}
\label{table:STOI-PESQ}
\end{table}

\section{Conclusion}
\label{sec:conclusion}
In this work, we proposed TFGAN, an improved MelGAN neural vocoder.
By the proposed time domain audio loss, the generator could capture waveform efficiently combining with STFT Loss.
What's more, we proposed the frequency discriminator to score the distance between real speech and generated speech.
Also, we improved the transpose convolution based upsampling mechanism by integrating nearest interpolation to reduce periodic artifacts.
The proposed neural vocoder has the same inference speed as MelGAN, but the fidelity of generated speech is much improved.
Our experimental results shown that TFGAN was slightly outperformed than WaveRNN, which is a strong autoregressive neural vocoder.

\vfill\pagebreak

\bibliographystyle{IEEEbib}
\bibliography{strings,refs}
\end{document}